# Digital divide among the States of Mexico:

# a comparison 2010-2020

Sergio R. Coria[1] and Luz M. García-García[2]

[1]**Institute of Informatics, University of the Sierra Sur**
**Miahuatlán de Porfirio Díaz, Oaxaca, México**
E-mails: coria@unsis.edu.mx, sergio.coria@gmail.com

[2] **Institute of Municipal Studies, University of the Sierra Sur**
**Miahuatlán de Porfirio Díaz, Oaxaca, México**
E-mail: luz2g@yahoo.com.mx

**Declarations of interest:** none.

This research did not receive any specific grant from funding agencies in the public, commercial, or not-for-profit sectors.

**Abstract**

Digital divide is one of the challenges that open government must face in mid- and low-income countries. In these contexts, inhabitants are left out of the benefits of information and communication technology (ICT), such as online government services. The present scenario that has emerged from the COVID-19 pandemic offers opportunities and challenges in ICT access and use to current and potential users all over the world. Therefore, it is important to know the advancement in digital inclusion in the recent years, particularly regarding the consumption and use of ICT goods and services. Thus, this article analyzes the Mexican case by comparing the availability of ICT in households of the states between the years 2010 and 2020. Open data from the Mexican Censuses of these two years are used to produce these analyses. The results suggest that inequalities prevail between South and Southeast states compared to Center and North states, and that cell telephone availability has increased, fostering Internet access.

1. **INTRODUCTION**

The Networked Readiness Index is a metrics proposed by the World Economic Forum (WEF) about the use of information and communication technologies (ICT) for development and competitiveness of countries. In 2009-2010, Mexico is ranked 78 out of 133. At comparing to other Latin American nations, it is ranked under Chile (40), Costa Rica (49), Uruguay (57), Colombia (60), and Brazil (61) [1]. In 2020, Mexico occupies ranking 63 out of 134; it is under Uru- guay (47), Chile (50), Costa Rica (54), and Brazil (59) [2]. Although Mexico advanced fifteen ranks in a decade, it is still under several (al- most the same) Latin American countries.

In turn, the Digital Development Dashboard [3], shows that 31% of population has basic skills to use ICT, 23% has standard skills, and 7% has advanced skills in Mexico (ITU, 2020). Interna- tional statistics suggest that Mexico is a country disadvantaged in ICT access and, therefore, a significant rate of its inhabitants is not familiar to ICT use.

In 2020, the COVID-19 pandemic was a major obstacle for the Mexican Institute for Statistics and Geography (INEGI) to carry out the population and housing census. Before the sanitary contingency, INEGI had planned the census fieldwork to take place from March 2 to 27, 2020 [4]. However, due to the pandemic, the Federal Government suspended non-essential government services from March 24. Then, Mexico entered sanitary emergency on March 30 and non-essential activities (both government and private) were also suspended. These activities included face-to-face censuses and surveys. Thus, censuses and surveys were rescheduled for June and July in different dates in each state [5].

Mexican statistics coincide with the international ones with respect to the digital lag of this country. For instance, the Survey on Availability and Use of Infor- mation Technology in Households (ENDUTIH 2020) [6] shows that there are 84,064,765 Internet users in Mexico, only 72.0% of the population aged six or older. Use of Internet is concentrated in 25-34 years-old population, corre- sponding to 19.1%; secondly, 18-24 years individuals (15.7 %). In turn, the demo- graphic sector that least uses Internet is of those with 55 years and older (10.4%). On the basis of INEGI's censuses of 2010 and 2020, the research problem is depicted below.

2. **PROBLEM DEFINITION**

This article analyzes the digital divide among the states of Mexico by comparing percentages of households owning ICT goods and services in 2010 and 2020. This comparison is timely and feasible as data of the 2020 Census is recently available.

Also, the comparison is valuable because the COVID-19 pandemic occurring since early 2020 in Mexico (before data recording of the 2020 Census) seems to introduce a relevant effect in ICT consumption and use in this country. Thus, analyzing ICT availability in households before and after the pandemic is highly interesting.

Since the 2010 Census, data on ICT in Mexican households is available as open data. This information shows that the South and Southeast states (*i.e.*, Guerrero, Oaxaca, and Chiapas) present the lowest rates of ICT goods and services. In contrast, Center and North states (*e.g.*, Ciudad de México, Nuevo León, Baja California, etc.) present the highest rates. In that year, Internet is the sixth ICT in households; personal computer (PC) is fifth, and landline telephone is fourth. Television is the first ICT, radio is second, and cell telephone is third. Other reason to analyze the digital divide phenomenon is that the Mexican government announced in 2018 the creation of CFE Telecomunicaciones e Internet para Todos, a government company promoting digital inclusion in low-income sectors of population.

The 2020 Census offers data on two relatively new ICT, namely videogame console and audio and video streaming, which can provide valuable insight about patterns on technology consumption. Based on the above, a series of interesting research questions are, for instance, ¿have the South and Southeast states increased the consumption of ICT since 2010? And ¿have the ICT reported by the 2010 Census presented any variation in their prevalence in households? In other words, ¿have Guerrero, Oaxaca and Chiapas risen in the inter-state ranking? And ¿have Internet, PC and cell telephone availabilities in households surpassed any other ICT between 2010 and 2020? These questions are answered below.

## 3. RELATED WORK

Digital divide is a concept that has been developed, approximately, since the early 1990s. At first, it was considered that the gap referred only to physical access to computers and the internet; but the concept evolved over time. Some authors consider different levels of the gap. The first level is almost always associated to physical access; the second level is associated to digital skills. [7] separate the divide into two aspects: the access divide (ethnicity, income, education, and age), and the divide in skills and capabilities (experience in the use of computers, general Internet use, online purchases, and information searches on the internet). [8] notes that a distinction should be made between an Internet access divide and a skills divide; the latter comprehends differences between groups of people in terms of skills necessary to effectively use the internet. A third level of digital divide is related to the use and usefulness of the internet, which is known as technological appropriation or the outcomes [9].

In addition, digital divide at the individual level is analyzed by a series of characteristics or determinants. Also [9] propose seven types of determinants:

sociodemographic, economic, social, cultural, personal, and motiva- tional. Sociodemographic determinants are, for instance, age, gender, and ethnic- ity. Economic determinants are income, family income, and family wealth. Social determinants are networking and political participation. Cultural capital and cul- tural possessions are identified in the cultural determinant. In turn, determinants such as free time and activities dedicated to health are found in the personal cate- gory. Motivational determinants are Internet use frequency, use variety, and use skills. Material determinants are internet access at home or the number of devices available.

Scientific literature identifies three aspects related to the digital divide at a micro level that represent individual users' capabilities: digital literacy skills (opera- tional, formal, integrational, strategic, and communicative), resources (computer access, Internet access), and variables related to people's sociodemographic char- acteristics, which are considered as determinants of internet skills: gender, age, ed- ucational level, and income [10, 11, 12, 13, 14]. Location and poverty concentration have been identified as important variables that contribute to the digital divide, too [15, 16].

Other studies mention that other factors different from sociodemographic varia- bles are also important. Some examples are civic mentality and information pov- erty, which could be related to too much or too little information or the fact that information is obtained from too few sources [17, 18]. [19] identify these components of the digital divide: 1) technical means, 2) autonomy of use, 3) models of use, 4) social support networks, and 5) skills.

In turn, [20] address digital divide at a macro level, in which fac- tors determining the gap at regional or national levels are studied. These and other authors identify four factors, as follows:

- Access to electricity, Internet, and devices, and quality of this access;
- Traditional and digital skills, including literacy, critical thinking, and entrepreneurship;
- Use of technology, public and private digital services, digital products and content, various types of work, social and civic engagement activities, as well as places of access to measure actual value creation and digital inclusion of marginalized communities; and
- A supportive environment, particularly in terms of affordability, legally valid identification, financial inclusion, trust, and security. [20]

Digital divide among regions is usually associated to urbanization issues and to differences between urban and rural areas. There is a direct relationship between urbanization and ICT. Whereas ICT promote the economy and social life, the pos- itive effects of access to and use of ICT vary due to differences among develop- ment levels. Developed areas enjoy the benefits of ICT; in contrast, underdevel- oped areas are left out [21]. This fact coincides with what is pointed out by other authors who have studied ICT availability on a regional basis. They state that areas with higher urbanization levels are also the ones with higher Internet access levels [22], involving that differences exist in Internet use per regions and that there are more users in urban areas.

## 4. METHOD

The method in this diachronic research is quantitative with a basic descriptive approach. Data of the 2010 and 2020 Population and Housing Censuses is obtained from the Mexican Institute for Statistics and Geography [23, 24]. Availability of six types of ICT in households is addressed in these anal- yses for years 2010 and 2020: Internet, personal computer (PC), landline tele- phone, cell telephone, television (TV), and radio. In addition, data about availabil- ity of videogame console, and audio and video streaming are analyzed for year 2020, only. For both years, the source data is obtained at a municipal level of ag- gregation consisting of absolute frequencies of households having any of the ICT goods and services. Then, the data is computed to produce absolute frequencies at state level. Finally, percentages of households per state are calculated.

## 5. RESULTS

The results are presented in a series of statistical charts and tables below. Figure 5.1 presents the national view of a comparison between years 2010 and 2020 of percentages of households with the most popular ICT goods and services in the states of Mexico. Table 5.1 presents detailed results corresponding to 2020. TV has been the ICT with highest percentages in both years (91.1 y 90.9%, respectively); these percentages have remained practically the same after ten years. Cell telephone is the second technology in households of the Mexican states; it has surpassed radio. The increase of cell telephone is 23.4 percentage points (from 64.0% in 2010 to 87.4% in 2020). Radio is the third ICT available in households in 2020; but interestingly, it shows a decrease of 10.7 percentage points (from 78.2 to 67.5%). Internet is the fourth ICT in households; it presents an interesting increase of 31.0 percentage points, the highest of all six technologies (from 21.0% to 52.0%). Personal computer (PC) is the fifth technology; it shows a relatively modest increase of 8.6 percentage points. At last, landline telephone is the sixth ICT; this presents a curious decrease of 5.1 points (from 42.5% a 37.4%). Each of these ICT is analyzed on a per-state basis below.

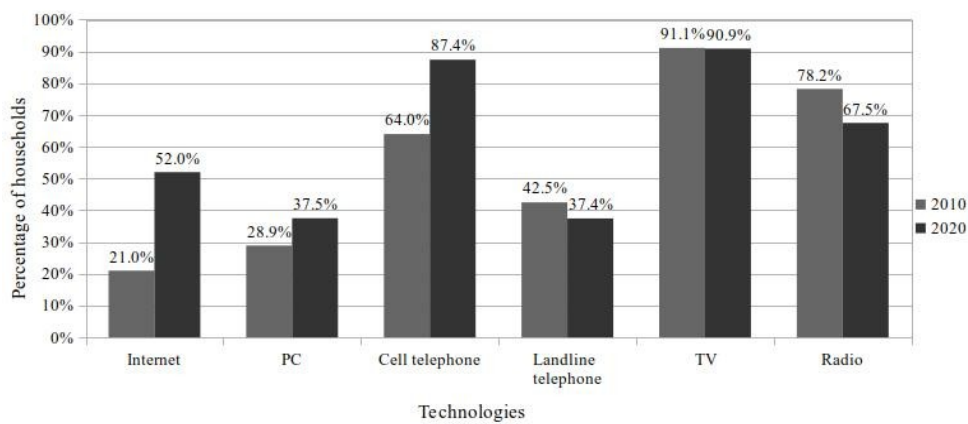

Figure 5.1. Comparison 2010-2020 of percentages of ICT in households: national view

Source: own elaboration using data from [23, 24]

### *5.1.* *Internet*

In 2020, Internet in households has increased in all the states of the country (Figure 5.2). Mexico City (CDMX) presents the highest availability percentage (75.6%); secondly, Baja California (BC, 69.6%), and third, Nuevo León (NL, 69.5%). The states with lowest percentages are Chiapas (Chis, 21.6%), Oaxaca (Oax, 29.3%), and Guerrero (Gro, 31.7%). The largest increases (as simple subtractions of percentage points, not as changes of proportion) occurred in Querétaro (Qro, 40.1 points), Aguascalientes (Ags, 38.3), and Nuevo León (NL, 38.3). The lowest increases occurred in Chis (14.6), Gro (20.9), and Oax (21.5).

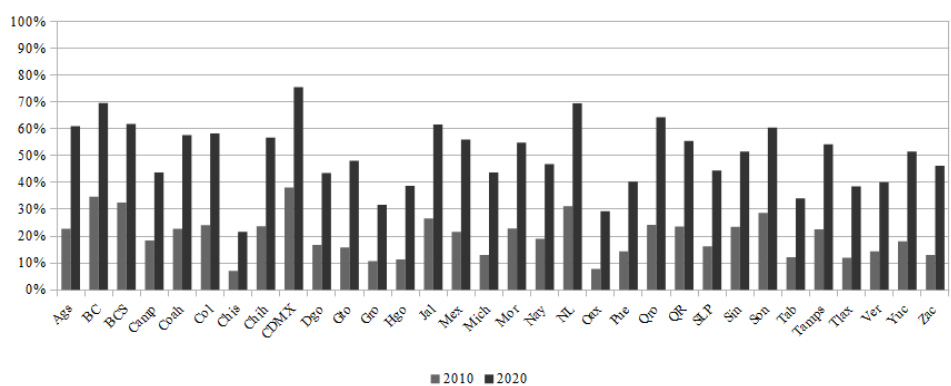

Figure 5.2. Percentages of households with Internet per state

Source: own elaboration using data from [23, 24]

| No. | State | Number of households | Internet | PC | Cell telephone | Landline telephone | TV | Radio | Videogame console | Streaming audio and video |
|---|---|---|---|---|---|---|---|---|---|---|
| 1 | Ags | 386,445 | 61.1% | 45.8% | 93.1% | 38.3% | 95.9% | 80.7% | 18.1% | 25.5% |
| 2 | BC | 1,148,913 | 69.6% | 50.2% | 94.0% | 50.2% | 93.3% | 64.4% | 18.9% | 33.4% |
| 3 | BCS | 240,468 | 61.8% | 45.3% | 94.2% | 38.2% | 89.1% | 59.4% | 15.1% | 28.3% |
| 4 | Camp | 260,725 | 43.7% | 33.7% | 83.7% | 26.4% | 86.7% | 52.3% | 6.9% | 14.8% |
| 5 | Coah | 900,883 | 57.7% | 40.8% | 91.5% | 41.3% | 95.9% | 71.3% | 13.8% | 21.7% |
| 6 | Col | 226,853 | 58.4% | 39.8% | 91.1% | 35.4% | 91.2% | 68.4% | 10.0% | 19.3% |
| 7 | Chis | 1,351,023 | 21.6% | 15.8% | 69.9% | 11.8% | 76.7% | 54.3% | 2.4% | 4.5% |
| 8 | Chih | 1,146,395 | 56.7% | 42.6% | 91.7% | 40.2% | 93.2% | 69.4% | 15.7% | 24.6% |
| 9 | CDMX | 2,756,319 | 75.6% | 59.8% | 92.0% | 68.9% | 96.1% | 78.6% | 20.6% | 34.7% |
| 10 | Dgo | 493,698 | 43.6% | 34.4% | 88.0% | 27.9% | 92.7% | 64.3% | 11.1% | 14.8% |
| 11 | Gto | 1,586,531 | 48.2% | 34.7% | 87.1% | 33.4% | 94.4% | 71.4% | 10.6% | 15.2% |
| 12 | Gro | 942,043 | 31.7% | 20.3% | 76.0% | 26.8% | 80.7% | 49.9% | 3.0% | 6.1% |
| 13 | Hgo | 857,174 | 38.8% | 30.5% | 84.9% | 24.6% | 88.6% | 67.7% | 7.5% | 11.2% |
| 14 | Jal | 2,330,706 | 61.7% | 44.4% | 91.6% | 43.3% | 94.5% | 72.8% | 16.1% | 24.6% |
| 15 | Mex | 4,568,635 | 56.1% | 40.5% | 88.6% | 46.5% | 94.3% | 73.9% | 12.5% | 18.0% |
| 16 | Mich | 1,284,644 | 43.7% | 29.1% | 87.4% | 25.9% | 91.8% | 66.7% | 7.9% | 11.7% |
| 17 | Mor | 560,669 | 54.9% | 36.5% | 89.1% | 42.2% | 90.9% | 69.5% | 8.6% | 16.2% |
| 18 | Nay | 361,270 | 46.9% | 34.8% | 88.2% | 29.9% | 89.8% | 62.4% | 10.0% | 16.7% |
| 19 | NL | 1,655,256 | 69.5% | 47.8% | 92.9% | 55.7% | 95.3% | 68.3% | 18.5% | 29.6% |
| 20 | Oax | 1,125,892 | 29.3% | 20.3% | 72.1% | 19.9% | 73.1% | 57.8% | 2.6% | 5.4% |
| 21 | Pue | 1,713,381 | 40.4% | 29.5% | 84.3% | 28.6% | 88.5% | 70.5% | 7.6% | 12.5% |
| 22 | Qro | 668,487 | 64.3% | 47.9% | 90.8% | 43.3% | 93.2% | 71.6% | 15.9% | 27.5% |
| 23 | QR | 575,489 | 55.5% | 37.2% | 91.4% | 27.6% | 86.0% | 57.6% | 9.7% | 22.9% |
| 24 | SLP | 774,658 | 44.5% | 34.0% | 84.3% | 34.5% | 89.7% | 67.1% | 10.8% | 16.6% |
| 25 | Sin | 854,816 | 51.6% | 38.2% | 92.1% | 24.6% | 93.1% | 57.4% | 10.8% | 19.0% |
| 26 | Son | 876,333 | 60.5% | 43.5% | 92.6% | 28.0% | 92.9% | 66.2% | 14.4% | 24.1% |
| 27 | Tab | 669,303 | 34.1% | 25.3% | 84.2% | 17.7% | 87.5% | 51.4% | 4.6% | 9.9% |
| 28 | Tamps | 1,069,121 | 54.3% | 35.6% | 92.1% | 38.6% | 93.1% | 61.1% | 10.9% | 18.4% |
| 29 | Tlax | 341,577 | 38.6% | 28.3% | 84.8% | 28.5% | 91.4% | 72.2% | 5.5% | 8.0% |
| 30 | Ver | 2,390,726 | 40.1% | 25.8% | 82.5% | 26.6% | 86.9% | 63.4% | 5.6% | 11.3% |
| 31 | Yuc | 658,085 | 51.5% | 37.4% | 88.2% | 26.8% | 91.1% | 66.0% | 9.1% | 19.6% |
| 32 | Zac | 442,623 | 46.3% | 31.5% | 84.2% | 31.2% | 94.0% | 71.1% | 8.5% | 11.7% |
|  | National | 35,219,141 | 52.0% | 37.5% | 87.4% | 37.4% | 90.9% | 67.5% | 11.5% | 18.8% |
|  | Max | 4,568,635 | 75.6% | 59.8% | 94.2% | 68.9% | 96.1% | 80.7% | 20.6% | 34.7% |
|  | Min | 226,853 | 21.6% | 15.8% | 69.9% | 11.8% | 73.1% | 49.9% | 2.4% | 4.5% |
|  | Avg. | 1,100,598.2 | 50.4% | 36.3% | 87.5% | 33.8% | 90.4% | 65.6% | 10.7% | 18.1% |

Table 5.1. Percentages of households with ICT per state in 2020

Source: own elaboration using data from [24]

## 5.2. *Personal computer (PC)*

Figure 5.3 shows that the states with highest availability of personal computer in 2020 are Ciudad de México (CDMX, 59.8%), Baja California (BC, 50.2%), and Querétaro (Qro, 47.9%). In contrast, the states with lowest are Chiapas (Chis, 15.8%), Guerrero (Gro, 20.3%) and Oaxaca (Oax, 20.3%). At comparing to 2010, the general increase in the country is relatively modest. The highest increases (as simple differences in percentage points, not as changes in proportion) occurred in Qro (14.0), CDMX (12.1), and Yucatán (Yuc, 11.8). The lowest increases occurred in Chis (3.4), Gro (4.4), and Tabasco (Tab, 4.7).

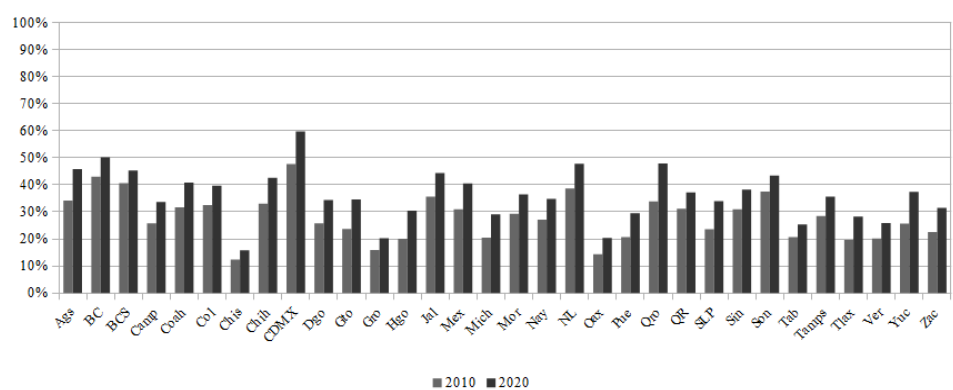

Figure 5.3. Percentages of households with PC per state

Source: own elaboration using data from [23, 24]

## 5.3. *Cell telephone*

Figure 5.4 presents the percentages of households with cell telephone per state. Interestingly, there are fourteen states with percentages over 90%. The states with highest availability are Baja California (BC, 94.2%), Baja California Sur (BCS, 94.0%), and Aguascalientes (Ags, 93.1%). Curiously, Mexico City (CDMX) is not the first ranked, but the eighth; although the differences in percentages are marginal, only. In the other hand, the lowest percentages correspond to Chiapas (Chis, 69.9%), Oaxaca (Oax, 72.1%), and Guerrero (Gro, 76.0%). The largest increases (as simple percentage subtraction) occurred in Puebla (Pue, 35.7), Oax (32.9), and Gro (32.6); whereas the lowest increases occurred in BCS (9.5), BC (11.7), and Quintana Roo (QR, 12.9).

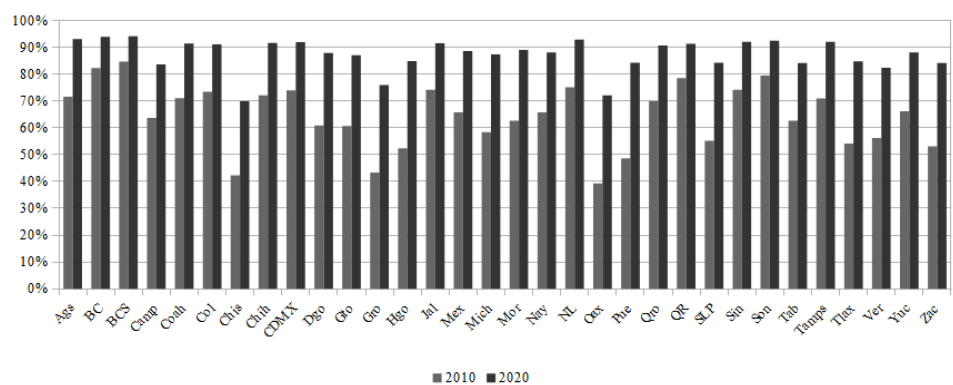

Figure 5.4. Percentages of households with cell telephone per state

Source: own elaboration using data from [23, 24]

### 5.4. *Landline telephone*

Landline telephone has presented a sensitive decrease of availability in households in 30 of 32 states from 2010 to 2020 (see Figure 5.5). Baja California (BC) and Querétaro (Qro) are the only states presenting a marginal increase (as simple percentage subtraction): 2.9 and 2.4 percentage points, respectively. The states with highest availability are Mexico City (CDMX, 68.9%), Nuevo León (NL, 55.7%), and Baja California (BC, 50.2%). In contrast, states with lowest availability are Chiapas (Chis, 11.8%), Tabasco (Tab, 17.7%), and Oaxaca (Oax, 19.9%). Usually, ICT rankings place Guerrero (Gro, 26.8%) among the three lowest states, but instead, it is ranked at 23 out of 32 in landline telephone availability in 2020. The largest decreases (as percentage subtraction) in this service occurred in Sinaloa (Sin, -17.8), Durango (Dgo, -15.4), and Sonora (Son, -14.3). The minimum dicreases occurred in Campeche (Camp, -0.3), Oax (-0.7), and Quintana Roo (QR, -1.0).

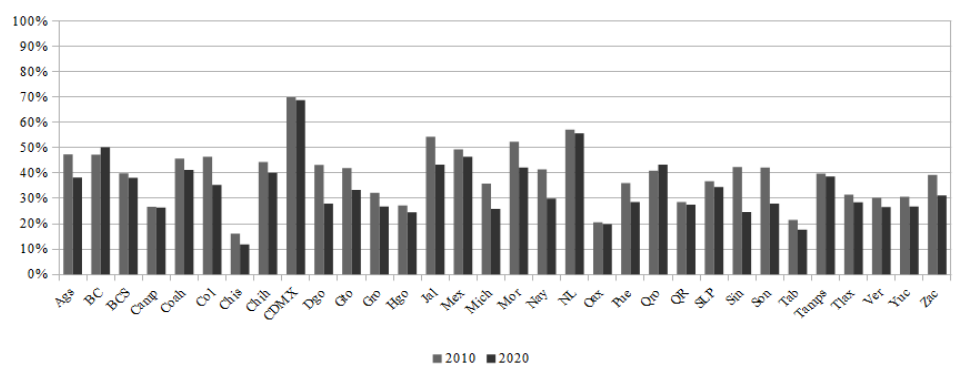

Figure 5.5. Percentages of households with landline telephone per state

Source: own elaboration using data from [23, 24]

## 5.5. Television (TV)

In general terms, percentages of households with TV remain the same after ten years (Figure 5.6) and variations seem to be marginal, only. The three maximum avail-ability percentages correspond to Mexico City (CDMX, 96.1%), Coahuila (Coah, 95.9%), and Aguascalientes (Ags, 95.9%). In the other hand, the three minimum correspond to Oaxaca (Oax, 73.1%), Chiapas (Chis, 76.7%) and Guerrero (Gro, 80.7%). The three highest marginal increases (as percentage subtractions) oc-curred in Chihuahua (Chih, 3.1), San Luis Potosí (SLP, 2.8), and Tamaulipas (Tamps, 2.4). In turn, the marginal decreases occurred in Baja California Sur (BCS, -3.5), Colima (Col, -2.8), and Morelos (Mor, -2.7).

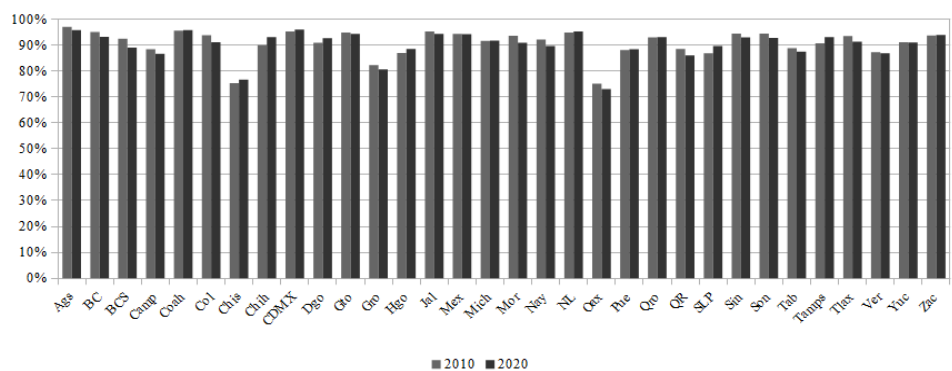

Figure 5.6. Percentages of households with TV per state

Source: own elaboration using data from [23, 24]

## 5.6. Radio

All states present a decrease in the presence of radio in households (Figure 5.7). The states with highest availability are Aguascalientes (Ags, 80.7%), Mexico City (CDMX, 78.6%), and State of Mexico (Mex, 73.9%). The states with minimum availability are Guerrero (Gro, 49.9%), Tabasco (Tab, 51.4%), and Campeche (Camp, 52.3%). Usually, Oaxaca (Oax, 57.8%) appears among the last three ranks in ICT; however, it is ranked at 26 out of 32. The three states with maximum de-creases (percentage subtractions) are: Baja California (BC, -18.0), Tab (-14.4), and Nuevo León (NL, -13.8). In turn, the minimum decreases occurred in: Puebla (Pue, -6.6), Ags (-7.6), and Michoacán (Mich, -7.8).

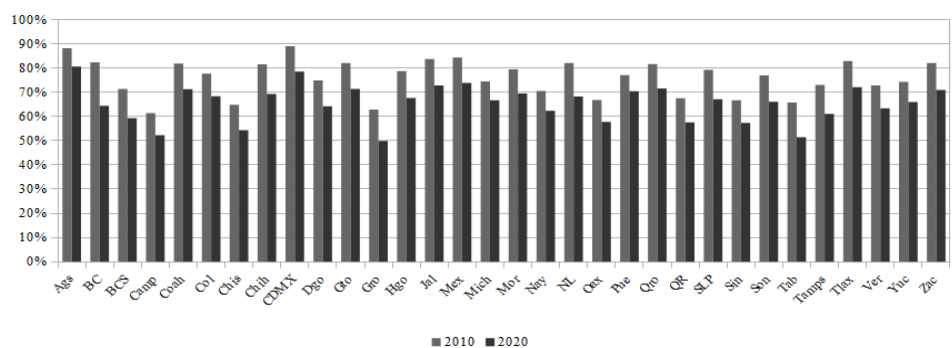

Figure 5.7. Percentages of households with radio per state

Source: own elaboration using data from [23, 24]

### 5.7. New ICT in 2020

The 2020 Mexican Census has incorporated into its database two new ICT types that are available in households nowadays, but were not available in 2010: specifically, videogame console (*e.g.*, PlayStation, Xbox, Nintendo, etc.), and streaming audio and video services (for instance, Netflix, AmazonPrime, Spotify, etc.). Table 5.1 presents percentages of households per state having eight ICT types, including the two new types abovementioned. Only 11.5% of households in the country own a videogame console. The three states with highest percentages are: Mexico City (CDMX, 20.6%), Baja California (BC, 18.9%), and Nuevo León (NL, 18.5%). In turn, the three states with lowest percentages are: Chiapas (Chis, 2.4%), Oaxaca (Oax, 2.6%) and Guerrero (Gro, 3.0%). Regarding streaming audio and video ser-vices, 18.8% of households in the country own them. The three states with maxi-mum rates are: Mexico City (CDMX, 34.7%), Baja California (BC, 33.4%), and Nuevo León (NL, 29.6%). Finally, the three states with minimum rates are: Chis (4.5%), Oax (5.4%), and Gro (6.1%).

### 6. DISCUSSION

In general terms, although Internet and cell telephone availability have increased from 2010 to 2020 in the whole country, the differences among South and Southeast compared to other regions are, still, large. Literature suggests that digital divide is greater for low-income regions; for instance, in rural regions, where there is a partial or total lack of digital infrastructure. This shows the consistency of our results to previous works. This is related to the first level of digital divide that is associated to ICT access and is basic to access digital skills or digital appropriation [9].

Besides, the states in South and Southeast Mexico present nowadays (and have presented for several decades) high levels of poverty and marginalization. This coincides to what has been pointed out by [15, 25, 16]; specifically, regarding that location and concentration of poverty are relevant variables contributing to digital divide.

The results are consistent to previous related work; for instance, to [7, 8, 12, 9, 20, 22], among others. Areas with higher urbanization and wealthy levels are also the ones with higher ICT availability and use. This is evident by contrasting the states in the Center, North and West of Mexico with those located at the South and Southeast.

A major implication of the results to open and digital government is that most of citizens in the states with largest digital disadvantages in 2020 (Guerrero, Oaxaca, and Chiapas) cannot benefit from digital government services as much as inhabitants of the other states in Mexico. However, as Internet and cell telephone availability have increased in the last ten years, there is an opportunity to exploit digital government services by means of the mobile government modality.

## 7. CONCLUSION

The results show that the South and Southeast states (Guerrero, Oaxaca, and Chiapas) increased the consumption and use of ICT from 2010 to 2020, but this increase does not suffice to reach the national averages. In general terms, these three states have stayed in the last three ranks since 2010; except that Guerrero had an interesting increase in landline telephone and it is located at 23 out of 32 in 2020. In turn, Oaxaca presents an increase in radio availability, and this state is located at rank 26 in this year.

Regarding the variation magnitudes of ICT availability, the largest increases (approximately, 39 percentage points in the top three states) correspond to Internet in Querétaro, Aguascalientes, and Nuevo León. Secondly, cell telephone increased mostly in Puebla, Oaxaca, and Guerrero (34 percentage points, approximately in these states). Personal computer (PC) presents a relatively modest increase, and their maximum ones correspond to Querétaro, Mexico City, and Yucatán (12.9 percentage points, approximately, in each). On the contrary, landline telephone presents a general decrease as it occurs in 30 out of 32 states: the largest decreases correspond to Sinaloa, Durango, and Sonora, with -16.0 percentage points (average) in these three states. Also, radio availability has decreased in Mexico; the largest decreases occurred in Baja California, Tabasco, and Nuevo León: -16.0 percentage points (average of these states). Finally, television (TV) remains with, practically, no variation since 2010.

Interesting phenomena can be identified; for instance, Internet availability has grown, but PC availability has not grown as much as Internet in the last ten years. Thus, it can be hypothesized that a relatively large percentage of Mexicans use Internet by means of other devices different from PC. These devices might be, for instance, cell telephone or tablet, although this hypothesis should be verified. Also, results suggest that landline telephone and radio functionalities might be being substituted by those of cell telephone. New ICT in households have appeared in the 2020 census: videogame console, and streaming video and audio services. Availability of these two technologies is still low in Mexico (11.5% and 18.8% of households, respectively), and the states in the South and Southeast are also the ones with lowest availability.

A basic public policy recommendation might be to reinforce the promotion of ICT in South and Southeast Mexico. Perhaps, plans and programs created by CFE Telecomunicaciones e Internet para Todos, the government company to supply Internet access to marginalized zones, should be fostered. Also, digital literacy and subventions should be promoted to increase the use of ICT goods and services in this region. Mobile government services can be potentially useful to its inhabitants.

This article has analyzed the variation of access to ICT goods and services in the states of Mexico solely on a descriptive and comparative basis by contrasting data from the 2010 and 2020 Censuses. This research work has analyzed statistical data recorded before the finish of the COVID-19 pandemic in Mexico, so the effects of this sanitary contingency on ICT consumption and use cannot be completely identified.

Future research work should develop qualitative investigation with an explanatory approach to know why digital divide among states in Mexico in 2020 remains similar to that in 2010. A relevant issue in this regard is the availability of ICT infrastructure, goods, and services in South and Southeast Mexico. This article has analyzed variables on digital divide at a regional level, which is focused on ICT access. Thus, these variables belong to the first level of digital divide. Therefore, variables in the micro level of digital divide should be analyzed (as literature recommends) to obtain further insight on this phenomenon. These variables are, for instance, ICT use, skills, and appropriation. As the pandemic forced and fostered the use of ICT for educational and labor purposes, further research is needed to analyze these effects. Finally, studies should be carried out on the effect that public policies implemented by CFE Telecomunicaciones e Internet para Todos can have on disadvantaged areas in the coming years.